\documentclass[twocolumn]{aastex61}

\received{xx xx, xxx}
\revised{xx xx, xxx}
\accepted{xx xx, xxx}

\shorttitle{Simulating the Fermi Bubbles as Forward Shocks}
\shortauthors{Zhang \& Guo}
\usepackage{graphicx}
\usepackage{subfigure}
\begin{document}
\bibliographystyle{aasjournal}

\title{Simulating the Fermi Bubbles as Forward Shocks Driven by AGN Jets}

\author{Ruiyu Zhang}
\affiliation{Key Laboratory for Research in Galaxies and Cosmology, Shanghai Astronomical Observatory, Chinese Academy of Science, 80 Nandan Road, Shanghai 200030, China}
\affiliation{University of Chinese Academy of Sciences, 19A Yuquan Road, Beijing 100049, China}
\author{Fulai Guo}
\affiliation{Key Laboratory for Research in Galaxies and Cosmology, Shanghai Astronomical Observatory, Chinese Academy of Science, 80 Nandan Road, Shanghai 200030, China}
\affiliation{University of Chinese Academy of Sciences, 19A Yuquan Road, Beijing 100049, China}

\correspondingauthor{Fulai Guo}
\email{fulai@shao.ac.cn}

\begin{abstract}

The Fermi bubbles are two giant bubbles in gamma rays lying above and below the Galactic center (GC). Despite numerous studies on the bubbles, their origin and emission mechanism remain elusive. Here we use a suite of hydrodynamic simulations to study the scenario where the cosmic rays (CRs) in the bubbles are mainly accelerated at the forward shocks driven by a pair of opposing jets from Sgr A*. We find that an active galactic nucleus (AGN) jet event happened $5-6$ Myr ago can naturally reproduce the bilobular morphology of the bubbles, and the postshock gas temperature in the bubbles is heated to $\sim0.4$ keV, consistent with recent X-ray observations. The forward shocks compress the hot halo gas, and at low latitudes, the compressed gas shows an X-shaped structure, naturally explaining the biconical X-ray structure in the ROSAT 1.5 keV map in both morphology and X-ray surface brightness. CR acceleration is most efficient in the head regions of the bubbles during the first 2 Myrs. The opposing jets release a total energy of $\sim 10^{55}$ erg with an Eddington ratio of $\sim 10^{-3}$, which falls well in the range of the hot accretion flow mode for black holes. Our simulations further show that the forward shocks driven by spherical winds at the GC typically produce bubbles with much wider bases than observed, and could not reproduce the biconical X-ray structure at low latitudes. This suggests that starburst or AGN winds are unlikely the origin of the bubbles in the shock scenario.

\end{abstract}


\section{Introduction} \label{sec:intro}

While searching for the gamma-ray counterpart of the WMAP haze at the Galactic center (GC) detected by the  {\it Wilkinson Microwave Anisotropy Probe} (WMAP) in microwave (\citealt{finkbeiner04a}; \citealt{dobler10}), two giant gamma-ray bubbles were discovered in 2010 in the \textit{Fermi}-LAT data, and have thereafter been referred as the ``Fermi Bubbles'' \citep{Su2010}. The bubbles have a bilobular shape, extending to $8-10$  kpc above and below the Galactic plane with sharp edges and approximately uniform surface brightness except for enhanced emissions from a ``cocoon'' structure located in the east part of the southern bubble (\citealt{Su2012,Ackermann2014}). At low latitudes, the bubble edges roughly coincide with the biconical X-ray structure previously discovered by \citet{bland03} in the ROSAT 1.5 keV map, suggesting that they may share the same origin \citep{bland19}. The bubbles have a unique spectrum $dN_{\gamma}/dE_{\gamma}\sim E_{\gamma}^{-2}$ between $1$ GeV and $100$ GeV with no spatial variations, which is significantly harder than other diffuse emission components in the inner Galaxy.

Although the Fermi bubbles have been discovered for many years, its emission mechanism and physical origin are still unclear. There are basically two types of emission mechanisms: hadronic or leptonic. In the hadronic scenario, inelastic collisions between cosmic ray (CR) protons and thermal nuclei produce pions, and then the neutral pions decay into gamma rays \citep{Crocker2011,Zubovas2011,Mou2014,Mou2015}. In the leptonic model, however, the gamma-ray emission is produced by the inverse Compton scattering between the ambient interstellar radiation field (ISRF) and CR electrons \citep{Guo2012a,Guo2012,Yang2017}. One major difference between the hadronic and leptonic models is the predicted age of the bubbles. In the hadronic model, there is essentially no upper limit on the bubble age (up to several billion years; \citealt{Crocker2011}), while in the leptonic scenario, due to the short cooling time of TeV electrons the age of the Fermi bubbles should be no more than several million years unless CR electrons are continuously accelerated in situ. Therefore, for models with different gamma-ray emission mechanisms, the energy source of the Fermi bubbles could be quite different.

Given the fact that the Fermi bubbles are roughly symmetric about the GC, it is quite often to relate their origin to active galactic nucleus (AGN) or nuclear starburst activities. Nuclear outflows driven by GC star formation activities have been proposed to explain the physical origin of the bubbles, and in these models, the gamma-ray emissions from the bubbles are hadronic with the predicted bubble age typically ranging from $\sim 25$ Myr to more than $100$ Myr \citep{Crocker2011,Crocker2015,Lacki2014,Sarkar2015}. It is also natural to associate the Fermi bubbles with recent activities of the supermassive black hole (SMBH) Sgr A* at the GC. With hydrodynamic simulations, \citet{Guo2012a} and \citet{Guo2012} show that an AGN jet event happened about $1-3$ Myr ago can naturally reproduce the Fermi bubbles with the observed location, size, and morphology (also see \citealt{guo17} for updated discussions). The AGN jet model adopts the leptonic scenario for the gamma-ray emission of the bubbles, and is further corroborated with three-dimensional (3D) magnetohydrodynamic (MHD) simulations by \citet{Yang2012}, \citet{yang13}, and \citet{Yang2017}. In addition to AGN jets, AGN winds from Sgr A* have also been suggested to power the Fermi bubbles, particularly by the hadronic model of \citet{Mou2014}, where the wind activity from Sgr A* lasted for about 10 million years and was quenched about 0.2 million years ago.

The physical conditions on the Fermi bubble edges are also unclear. The sharp bubble edge has been inferred as a termination shock \citep{Lacki2014}, a forward shock \citep{Fujita2013}, or a contact discontinuity \citep{Crocker2011,Guo2012a, Mou2014, Sarkar2015}. More recently, \citet{Keshet2017} analyze the 8-year Fermi-LAT data, and their result is in favor of the bubble edge as a forward shock. While most models discussed above assume that CRs are accelerated at the GC and then transported to the current bubble location by either jets or winds, CRs can also be accelerated in situ by shocks \citep{Keshet2017,Cheng2011,Cheng2015,Fujita2013,Fujita2014,Lacki2014} or turbulence \citep{Mertsch2011,Mertsch2019,Cheng2014,Sasaki2015}. For a detailed comparison between these models, the readers are referred to \citet{Yang2018} for a review.

Shocks are natural accelerators of CR particles. In 1949,  Enrico Fermi first proposed a mechanism with which particles can statistically gain energy through repeatedly collisions with clouds in the interstellar medium \citep{Fermi1949}. In the late 1970s, this mechanism was further shown to be particularly efficient at shock fronts, and was called diffusive shock acceleration (DSA; see \citealt{Axford1977,Krymskii1977,Bell1978,Blandford1978}). DSA naturally results in a power law energy spectrum ($E^{-2}$), with  $\sim10 \%$ energy in the shock transferred to CRs. In supernova remnants, it is thought that DSA is the most possible mechanism that accelerates particles up to $10^{15}$ eV. Similarly, this mechanism may also be responsible for acceleration of the CRs in the Fermi bubbles, as suggested in \citet{Fujita2013}, which show that the observed flat gamma-ray surface brightness and sharp edges of the bubbles can be reproduced in a simple spherically-symmetric shock model. \citet{Fujita2014} further show that the gamma-ray spectrum predicted in their model is consistent with the observations, although they could not distinguish whether the shocks are created through starbursts or AGN activities at the GC.

Studies on the properties of the hot gas in the Galactic halo and the Fermi bubbles suggest that the edge of the bubbles may indeed be shock fronts. \citet{Kataoka2015} studied the \textit{Suzaku}  and \textit{Swift} data and found that gas temperatures inside both the northern and southern bubbles are almost constant $k_{\rm B}T \sim 0.30 \pm 0.07 $ keV, while the gas temperature in the Galactic halo is known to be $k_{\rm B}T\sim0.2$ keV \citep{Kataoka2018}. This suggests that the gas in the bubbles could be heated by a weak shock with Mach number $M \sim 1.5$. Similarly, studies on the O VII and O VIII line strengths by \citet{Miller2016} suggest that the gas temperature in the bubbles is $k_{\rm B}T \sim 0.40 $ keV, corresponding to a shock with Mach number $M \simeq 2$ at the bubble edge. In a continuous injection model, \citet{Miller2016} further estimate the age of the bubbles to be $3-5$ Myr with an energy injection rate of $\sim(1-7) \times 10^{42}$ erg s$^{-1}$.

Additional clues came from studies of the gas kinematics inside the Fermi bubbles with ultraviolet absorption-line spectra.
\citet{Bordoloi2017} studied the spectra of 47 background AGNs and halo stars in and around the northern bubble, and found that the velocities of the observed blueshifted absorption components range from 265 km s$^{-1}$ to 91 km s$^{-1}$ in the Galactic standard of rest (GSR) coordinate system, decreasing with Galactic latitude (also see \citealt{Karim2018}). This could be explained by a nuclear outflow happened about $6-9$ Myr ago, while a similar study by \citet{fox15} suggests that the outflow age is $\sim 2.5-4$ Myr. These age estimates are roughly consistent with the bubble age in the shock model inferred from X-ray observations described in the previous paragraph. We also note that the bipolar radio bubbles recently discovered in the GC may be a less energetic version of the Fermi bubble event which also happened several million years ago \citep{Heywood2019}.

Combining the constraints on the age of the Fermi bubbles from X-ray and ultraviolet observations, we may take the bubble age to be $\sim 5 $ Myr, which is longer than that inferred in leptonic models (1--3 Myr), but shorter than that inferred in hadronic wind models ($>10$ Myr). The \textit{in situ} shock acceleration model shows some advantage to address this discrepancy. If the CR particles in the bubbles are accelerated by the shock rather than transported from the GC, then there will be essentially no age constraints on the bubbles due to the short cooling time of CR electrons, and the bubble age only depends on the dynamical process that created the bubbles.

In this work, we explore the shock model of the Fermi bubbles with hydrodynamic simulations. Assuming that the CRs within the bubbles are mainly accelerated by the forward shock at the expanding bubble edge, we use hydrodynamic simulations to study the dynamical evolution of the Fermi bubbles and compare our results with observations. While the shocks could be driven by AGN jets, AGN winds or nuclear star formation winds, CR particles can be accelerated more efficiently in strong shocks \citep{Fujita2013}. Therefore we focus on the AGN jet model, and investigate whether some main observed properties of the bubbles, e.g., the bilobular bubble morphology, can be reproduced in our simulations. We also compare our results with X-ray observations, including the ROSAT 1.5 keV X-ray map \citep{Snowden1997} and the gas temperatures in the bubbles recently measured by \citet{Kataoka2015} and \citet{Miller2016}. The rest of the paper is organized as follows. In Section \ref{sec2}, we describe the details in our simulations, including our model for the Galaxy and numerical setup. We present the results of a representative run in Section 3 and investigate the impacts of several important model parameters in Section 4. We summarize the results in the last section.

\section{Numerical Methods} \label{sec2}

Assuming axisymmetry around the rotational axis of the Milky Way, we investigate the formation of the Fermi bubbles with hydrodynamic simulations. We use the same hydrodynamic code previously used in \citet{Guo2012a} and \citet{Guo2012}. It is a 2-dimensional (2D) axisymmetric Eulerian grid-based code using algorithms similar to ZEUS 2D \citep{stone1992}. The code has also been successfully used to study AGN feedback in galaxy clusters (e.g., \citealt{guo11}; \citealt{guo18}). Here in this section we present the model assumptions and simulation setup, with a focus on several improvements. In particular, we adopt an updated model from \citet{McMillan2017} for the mass distribution of the Milky Way, which includes contributions from the dark matter halo, the Galactic bulge, the thin and thick stellar disks, and the HI and molecular gas disks.

\subsection{Basic Equations}

The following equations are solved in our simulations:
\begin{eqnarray}
  \frac{d\rho}{dt}+\rho\nabla\cdot\textbf{v} &=& 0 \\
   \rho\frac{d\textbf{v}}{dt}+\nabla(P+P_{c})&=&-\rho\nabla\Phi \\
   \frac{\partial e}{\partial t}+\nabla\cdot(e\textbf{v})&=& -P\nabla\cdot\textbf{v} \\
 \frac{\partial e_{c}}{\partial t}+\nabla\cdot(e_{c}\textbf{v})&=& -P_{c}\nabla\cdot\textbf{v}+\nabla\cdot(\kappa\nabla e_{c})
\end{eqnarray}
where $\rho, {\bf v}, P, e$, are the  density, velocity, pressure and energy density of thermal gas, respectively. As in \citet{Guo2012a}, CRs are treated as a second fluid. $P_{c}$ and $e_{c}$ are the pressure and energy density of CRs. As usual, we assume that $P=(\gamma-1)e$ and $P_{c}=(\gamma_{c}-1)e_{c}$, where $\gamma=5/3$ and $\gamma_{c}=4/3$ are the adiabatic indices of thermal gas and CRs, respectively. $\kappa$ is the CR diffusion coefficient, and we set $\kappa=3\times10^{27}$ cm s$^{-1}$ in the simulations (see \citealt{Guo2012a} for relevant discussions on $\kappa$).

We assume that the hot gas follows the ideal gas law:
\begin{equation}\label{eq0}
  T=\frac{\mu m_{\mu}P}{k_{\rm B}\rho}{~,}
\end{equation}
where $k_{\rm B}$ is Boltzmann's constant, $\mu$ = 0.61 is the mean molecular weight per particle, $m_{\mu}$ is the atomic mass unit, and T refers to the temperature of the gas. The relationship between the gas density and the electron number density $n_{\rm e}$ is
\begin{equation}\label{eq-rou}
\rho=\mu_{\rm e}n_{\rm e}m_{\mu}{~,}
\end{equation}
where  $\mu_{\rm e}=5\mu/(2+\mu)$ is the molecular weight per electron.

For simplicity, magnetic fields are not included in our simulations. To reproduce the observed morphology of the Fermi bubbles, the jets in our simulations are kinetic-energy-dominated. Magnetic fields are dynamically unimportant in the evolutions of the jet ejecta and the forward shocks (\citealt{Yang2017}).

\subsection{The Milky Way Model} \label{sec21}

To determine the Galactic potential well $\Phi$, we adopt the mass model of the Milky Way proposed in \citet{McMillan2017}. In this model, the Milky Way's mass is contributed by six static axisymmetric components:  the Galactic bulge, the thin and thick stellar disks, the HI and molecular gas disks, and the dark matter halo.

The density distribution of the Galactic bulge is
\begin{equation}\label{eq2}
\rho_{b}=\frac{\rho_{0,b}}{(1+r^{\prime}/r_{0})^{\alpha}}\textrm{exp}[-(r^{\prime}/r_{\rm cut})^{2}],
\end{equation}
where the scale density $\rho_{0,b}=9.93\times10^{10}$ M$_{\odot}$ kpc$^{-3}$,  $r_{0}=0.075$  kpc, $r_{\rm cut}=2.1$  kpc, $\alpha$=1.8, and in cylindrical coordinates, $r^{\prime}=\sqrt{R^{2}+(z/q)^{2}}$ with the axis ratio $q=0.5$. The corresponding total bulge mass is $M_{\rm b}=9.23 \times 10^{9}$ M$_{\odot}$.

The stellar disk consists of two parts: the thin and thick disks. Their density distributions can be depicted in the following form:
\begin{equation}\label{eq3}
\rho_{d}(R,z)=\frac{\Sigma_{0}}{2z_{d}}\textrm{exp}\left(-\frac{|z|}{z_{d}}-\frac{R}{R_{d}}\right),
\end{equation}
where $\Sigma_{0}$ is the central surface density, $z_{d}$ is the scale height, and $R_{d}$ is the scale length. For the thick disk,
 $\Sigma_{0}=183 \textrm{ M}_{\odot} \textrm{ pc}^{-2}$,  $z_{d}=900$  pc,  $R_{d}=3.02 \textrm{ kpc}$. For the thin disk, $\Sigma_{0}$=896 M$_{\odot} \textrm{ pc}^{-2}, z_{d}=300\textrm{ pc}, R_{d}=2.50$ kpc.

The gas disk also includes two components: the HI disk and the molecular gas disk. Each of these two gas disks has the following density distribution:
\begin{equation}\label{eq4}
\rho_{d}(R,z)=\frac{\Sigma_{0}}{4z_{d}}\textrm{exp}\left(-\frac{R_{m}}{R}-R/R_{d}\right)\textrm{sech}^{2}(z/2z_{d}),
\end{equation}
where  $\Sigma_{0}=53.1 \textrm{ M}_{\odot}\textrm{ pc}^{-2}$,  $z_{d}=85$  pc, $R_{d}=7$   kpc and $R_{m}=4$   kpc for the HI disk. For the molecular gas disk,  $\Sigma_{0}$ = 2180 $\textrm{M}_{\odot}\textrm{ pc}^{-2}$, $z_{d}=45$  pc, $R_{d}=1.5 $ kpc and  $R_{m}=12$ kpc.

The density distribution of the dark matter halo is described by the Navarro-Frenk-White (NFW) profile \citep{Navarro1996}
\begin{equation}\label{eq5}
\rho_{h}=\frac{\rho_{0,h}}{x(1+x)^{2}},
\end{equation}
where $\rho_{0,h}=0.00854$ $  \textrm{M}_{\odot} $ pc$^{-3}$ and $x=r/r_{h}$ with the scale radius $r_{h}=19.6$ kpc. Here $r=\sqrt{R^{2}+z^{2}}$ is the distance to the GC. In this model, the virial mass of the dark matter halo is $M_{\rm v}=1.37 \times 10^{12}$ M$_{\odot}$ with the concentration $c_{\rm v}=15.4$ \citep{McMillan2017}.

\begin{figure}
  \centering
  \includegraphics[width=8cm]{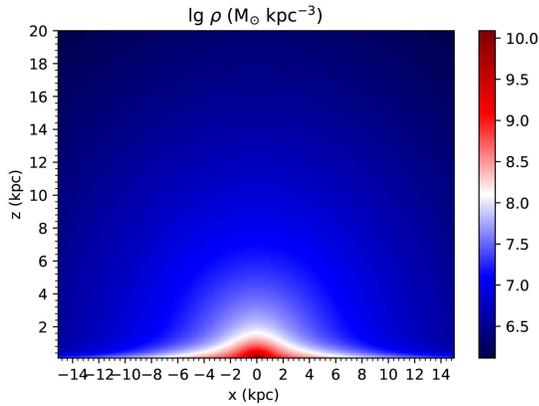}\\
  \caption{Density distribution of the Milky Way. Six axisymmetric components are included:  the Galactic bulge, the thin and thick stellar disks, the HI and molecular gas disks, and the NFW dark-matter halo.}\label{fig1}
\end{figure}

\begin{figure}
  \centering
  \includegraphics[width=8cm]{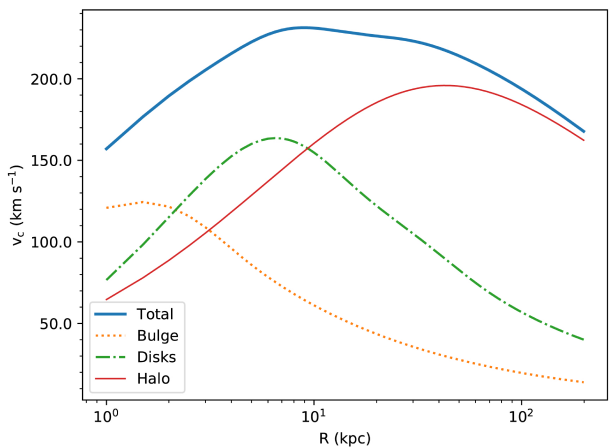}\\
  \caption{Rotation curve along the Galactic plane for the mass model adopted in this paper (see Sec. 2.2). The dotted, dash-dotted, and thin solid lines represent the contributions from the bulge, all the stellar and gas disks, and the dark matter halo, respectively.
   }\label{fig-rk}
\end{figure}

Combining Equations (\ref{eq2}--\ref{eq5}), we obtain the mass density distribution of the Milky Way. As shown in Figure \ref{fig1}, the disk-like morphology is well reproduced by this mass model. The gravitational potential of the Galaxy can be further calculated by solving the Poisson's equation. The rotational curve in the resulting Galactic potential along the $R$ direction (the Galactic plane) is shown in Figure \ref{fig-rk}, which shows good consistency with Figure 5 in \cite{McMillan2011}.

\subsection{Simulation Setup and Initial Conditions}

Equations (1) - (4) are solved in cylindrical coordinates $(R, z)$ assuming axisymmetry around the Galactic rotational axis. The computational domain along each axis consists of 1800 equally spaced grids in the inner 15 kpc and 100 logarithmically spaced zones from 15 to 70 kpc. The corresponding spatial resolution in the inner 15 kpc is $8.33$ pc. Along both the $R$ and $z$ directions, we adopt reflective boundary conditions at the inner boundaries and outflow boundary conditions at the outer boundaries.

For initial conditions at $t=0$, we assume that the hot gas is in hydrostatic equilibrium in the Galactic potential well, and there are no CRs in the Galaxy. Recent X-ray observations indicate that the temperature of the hot gas in the Galactic halo is around 0.2 keV, with little variations across different lines of sight \citep{Henley2013, Miller2015, Kataoka2018}. Therefore, among all the simulations in this work we assume that the hot gas is initially isothermal with a constant temperature $T = 2.32\times 10^{6}$ K ($\sim 0.2$ keV). Given the value of the gas temperature, the initial density distribution of the hot plasma can be derived from hydrostatic equilibrium. The normalization of the density distribution is determined by the initial electron number density at the origin $n_{e0}$.

\begin{figure}[h]
  \centering
  \includegraphics[width=8cm]{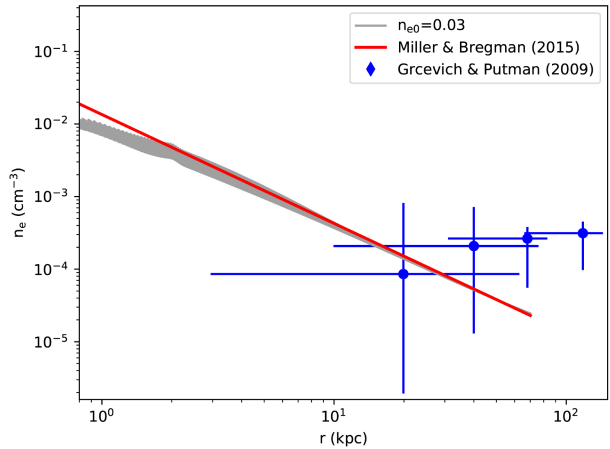}\\
  \caption{Initial profile of thermal electron number density as a function of radius in our fiducial run (run A). The grey-shaded area represents our model derived from hydrostatic equilibrium and a uniform temperature $T = 2.32\times 10^{6}$ K. At $r\gtrsim 2$ kpc, our model agrees very well with the best-fit $\beta$ model presented in \citet{Miller2015}. For comparison, the blue points show the gas densities derived in \cite{Grcevich2009}.}\label{fig-miller}
\end{figure}

In the fiducial run (run A), we choose $n_{e0}=0.03$ cm$^{-3}$, and the resulting thermal electron number density distribution is shown as the grey-shaded area in Figure \ref{fig-miller}. Note that the gas distribution is not spherically symmetric due to the non-spherically symmetric gravitational potential well. We compare the gas density distribution in our model with the $\beta$ model (the red solid line in Fig. \ref{fig-miller}) in \citet{Miller2015} derived from O VII and O VIII observations. At $r\gg r_{c}$, the best-fit $\beta$ model can be  described as \citep{Miller2015}:
  \begin{equation}\label{beta}
    n(r)\approx\frac{n_{0}r_{c}^{3\beta}}{r^{3\beta}}
  \end{equation}
 where $n_{0}r_{c}^{3\beta}$ = 0.0135 cm$^{-3}$ kpc$^{3\beta}$ and $\beta=0.5$.

It is remarkable that our initial gas density distribution agrees very well with the best-fit $\beta$ model presented in \citet{Miller2015} at $r\gtrsim 2$ kpc. As shown in Figure \ref{fig-miller}, our model also roughly agrees with the gas densities at $r\sim 10 - 50$ kpc measured by \citet{Grcevich2009} using the ram-pressure stripping argument of several Local Group dwarf galaxies. We note that beyond $\sim 70$ kpc, the density profiles in our model and the $\beta$ model lie substantially below the data points in \citet{Grcevich2009}. However, in the current work we focus on the Fermi bubble event and gas dynamics within the inner halo ($r\lesssim 10$ kpc), and the outer boundary of our simulations is set to be 70 kpc. Therefore, the gas distribution in the outer halo beyond 70 kpc will not affect our results.

From an observational point of view, the density distribution of the hot Galactic halo gas is quite uncertain (\citealt{Bregman2018}), and there are many density models proposed in the literature \citep[e.g.,][]{Maller2004,Yao2009,Fang2013,Miller2015,Nakashima2018,fang20}. The halo gas density distribution would affect the morphology of the resulting forward shock and the properties of the postshock gas \citep{Sofue2019}. Our model is based on hydrostatic equilibrium, and reproduces reasonably well the observed biconical X-ray structure near the GC in both morphology and X-ray surface brightness (Section 3.2), and the emission measures along many lines of sight toward the Fermi bubbles (Section 3.3).

\subsection{Jet  Setup}

We adopt the same jet setup as in \citet{Guo2012a}, where the jet at the base is hybrid, consisting of both thermal gas and CRs. In the current paper we focus on the hydrodynamic evolution of AGN jets and the resulting Fermi bubbles, and there are no CRs in the halo except those injected in the jets. As discussed in \cite{Guo2012a}, the jet parameters are highly degenerate. We have run a large suite of simulations, and in this paper, we present the results of several representative runs, as listed in Table \ref{tab-para}. Run A is the fiducial run, which matches the observational results best among all the simulations. In run B, we choose a much larger value for $n_{e0}$ to investigate the impact and uncertainties of the initial gas densities in the halo. Run C is performed to match the observed gas temperature in the Fermi bubbles presented in \cite{Kataoka2013}. In run D, we investigate the formation of the Fermi bubbles from a spherical outflow from the GC.

As in \citet{Guo2012a}, the jet in each simulation is launched along the $z$ axis at $t=0$ with a constant speed $v_{j}$ for a duration of $t_{j}$. The initial jet is implemented in a cylinder with radius $R_{j}$ and height $z_{j}$ along the $z$ axis starting from the GC. At the base, the jet contains both thermal gas with density $\rho_{j}$ and energy density $e_{j}$ and CRs with energy density $e_{\rm jcr}$. We define the jet density contrast $\eta$ at the jet base as the ratio between $\rho_{j}$ and the ambient gas density $\rho_{\rm amb}$.
As shown in \citet{guo15}, the shape evolutions of the ejecta bubble and the forward shock are mainly determined by two jet parameters: $\eta$ and the ratio of its kinetic energy density to non-kinetic energy density, and the latter includes both thermal and CR energy densities. $e_{j}$ and $e_{\rm jcr}$ are thus degenerate in our hydrodynamic simulations \citep{Guo2012a}. In our jet simulations, we set $e_{j}$ to be equal to the ambient gas energy density, while leaving $e_{\rm jcr}$ as a free parameter to fit the observed morphology of the Fermi bubbles. In our spherical wind simulations, we follow previous wind simulations (e.g., \citealt{Mou2014,Sarkar2017}) to set $e_{\rm jcr}=0$ and use $e_{j}$ as a free parameter.
We set $z_{j}=0.35$ kpc to avoid simulating the complex environment around the GC. We further assume that the jet has undergone significant deceleration within the jet initialization zone.

The parameters in our simulations are listed in Table \ref{tab-para}. In each row, $t_{\rm bub}$ refers to the current age of the Fermi bubbles, and $P_{\rm j}$ and $E_{\rm j}$ are respectively the power and total injected energy of one jet in the corresponding run. Note that in run D, we initialize a spherical wind instead of a collimated jet, and here $R_{j}$ and $z_{j}$ stand for the radius of the central spherical region used to set up the wind. $v_{j}$, $\rho_{j}$, $e_{j}$, and $e_{\rm jcr}$ represent the radial velocity, thermal gas density, thermal energy density, and CR energy density of the wind at its base, respectively. $t_{j}$ is the duration of the wind.

\begin{deluxetable*}{cccccccccccc}
\tablecaption{List of Our Simulations with Model Parameters and Some Key Results \label{tab-para}}
\tablewidth{0pt}
\tablehead{
\colhead{Run}  & \colhead{$n_{e0}$} &$\rho_{j}$&$e_{j}$&\colhead{$v_{j}$}   &\colhead{$e_{\rm jcr}$}  &  \colhead{$R_{j}$} & \colhead{$z_{j}$}& \colhead{$P_{\rm j}$}      & \colhead{$t_{j}$}           & \colhead{$t_{\rm bub}$}   & \colhead{$E_{\rm j}$}
 \\
\colhead{ID}            &\colhead{cm$^{-3}$}&  g cm$^{-3}$  &erg cm$^{-3}$       &\colhead{$10^{9}$cm $s^{-1}$}   & \colhead{erg cm$^{-3}$} & \colhead{pc }   & \colhead{pc }   & \colhead{erg~s$^{-1}$}      &   \colhead{Myr}   &   \colhead{Myr}      &   \colhead{erg}
}
\startdata
A         &0.03  &$1.23\times10^{-27}$ & $1.46\times10^{-11}$    &$2.5$   &$2.7\times10^{-10}$   & 33.3  & 350    &$3.42\times10^{41}$ & 1.0  & 5&$1.07\times10^{55}$\\
B         &0.3   &$1.23\times10^{-26}$ &$1.46\times10^{-10}$     &$2.5$    &$2.7\times10^{-9}$   & 33.3  & 350&   $3.42\times10^{42}$ & 1.0   & 5&$1.07\times10^{56}$\\
C        &0.03   &$1.23\times10^{-27}$ &$1.46\times10^{-11}$    &$2.1$    &$2.0\times10^{-10}$   & 25.0  & 350         &$1.15\times10^{41}$  & 1.0 & 6&$3.61\times10^{54}$\\
D        &0.03   & $2.27\times10^{-27}$& $2.68\times10^{-11}$   &$0.9$    &$0$      & 41.6  & 41.6       &$8.79\times10^{40}$ & 5.0  &9& $1.38\times10^{55}$\\
\enddata
\tablecomments{In our simulations, the jet is implemented in a cylinder with radius $R_{j}$ and height $z_{j}$ along the $z$ axis. At its base, the jet is parameterized with five parameters: gas density $\rho_{j}$, thermal energy density $e_{j}$, CR energy density $e_{\rm jcr}$, velocity $v_{j}$, and duration $t_{j}$. $P_{\rm inj}$ and $E_{\rm inj}$ refer to the power and the total injected energy of one jet, respectively. $t_{\rm bub}$ is the current age of the Fermi bubbles in each simulation. $n_{e0}$ is the initial electron number density at the origin, which determines the normalization of the initial density distribution in the halo. Run D is a spherical wind simulation investigated in Section 4.3, and here $R_{j}$ and $z_{j}$ stand for the radius of the central spherical region used to set up the wind. $v_{j}$ in run D refers to the radial velocity, instead of the z-component velocity as in jet simulations.}
\end{deluxetable*}

\section{The Shock Model: A REPRESENTATIVE RUN} \label{sec-res}

In this section, we present the main results of our fiducial run (run A), which matches the observations best. Two constraints are used to determine the best-fit simulation. The first constraint is the morphology of the resulting bubble. In the shock model, we assume that the forward shock driven by the AGN jet event represents the edge of the Fermi bubbles. By adjusting the parameters of the jet, the location, size and bilobular morphology of the Fermi bubbles can be reproduced. If the jet is very powerful, a strong shock will be created, and it then takes less time to reach the current size of the bubbles. On the other hand, a weak shock results in a longer age of the current Fermi bubbles. To further constrain the jet power and the age of the bubbles, we use the second constraint -- the temperature of the shock-compressed gas in the Fermi bubbles. In run A, we use the gas temperature $\sim 0.4$ keV measured by \citet{Miller2016}. The age and energy of the Fermi bubbles can be largely determined by these two constraints. We stop the simulation when the forward shock roughly reaches the edge of the Fermi bubbles.

The jet parameters in run A is listed in Table 1. Both the northern and southern Fermi bubbles show quite narrow bases near the GC. To reproduce this feature in the shock model, the initial jet radius must also be quite small, and in run A we choose R$_{j}=33.3$ pc, about one order of magnitude smaller than that adopted in \citet{Guo2012a}. The jet in run A is light, with a density contrast $\eta=0.04$ compared to the ambient gas. The jet is kinetic-energy-dominated and the kinetic power accounts for about $93\%$ of its total power. At the jet base, the values of $e_{j}$ and $e_{\rm jcr}$ are highly degenerate, and the total pressure in the jet affects the shape of the resulting bubble (forward shock). During its active phase, the jet has a total power of $3.42\times10^{41}$ erg s$^{-1}$, and with a duration of 1 Myr, the total injected energy is $1.08\times10^{55}$ erg. Taking a jet feedback efficiency of 10\%, the mass accretion rate of the central supermassive black hole Sgr A* can be estimated $\dot{M}_{\rm BH}=2P_{\rm j}/(0.1c^{2})=1.2\times 10^{-4}$ $M_{\sun}/$yr, and the total mass accreted by Sgr A* during this event is $120$ $M_{\sun}$. During the active phase, the Eddington ratio of Sgr A* is $\epsilon=2P_{\rm j}/L_{\rm Edd}\sim1.2\times 10^{-3}$, which falls well in the range of the hot accretion flow mode for SMBHs \citep{Yuan2014}.

In the remainder of this section, we will first show the morphology, gas density, velocity and temperature of the Fermi bubbles in the simulation, and compare them with observational values. Then, we will present the synthetic X-ray surface brightness map in our simulation and show that the X-shaped biconical structure in the ROSAT 1.5 keV map is reproduced in our simulation. We calculate the emission measures along the lines of sight toward the Fermi bubbles in the simulation, and compare them directly with the observational data. At last, we will present the evolution of the Mach number at the outer edge of the simulated bubble. The observed Fermi bubble morphology is a 3D structure projected onto the 2D sky map. With axisymmetry around the Galactic rotational axis, our cylindrical coordinates $(R, z)$ centered at the GC, can be naturally converted to the Cartesian coordinates $(x, y, z)$, which is connected with the Galactic coordinates $(l, b)$ centered at the solar system through
\begin{equation}\label{proj}
\textrm{tan}l=-\frac{x}{y+R_{\odot}}
\end{equation}
\begin{equation}\label{proj2}
\textrm{tan}b=\frac{z}{\sqrt{x^2+(y+R_{\odot})^{2}}}
\end{equation}
where the location of the Sun is (0, $-R_{\odot}$, 0) in the Cartesian coordinates, and here we set $R_{\odot} = 8.5$ kpc.

\subsection{Properties and Evolution of the Bubble}

\begin{figure}[h!]
 \subfigure[]{
 \centering
   \includegraphics[width=8cm]{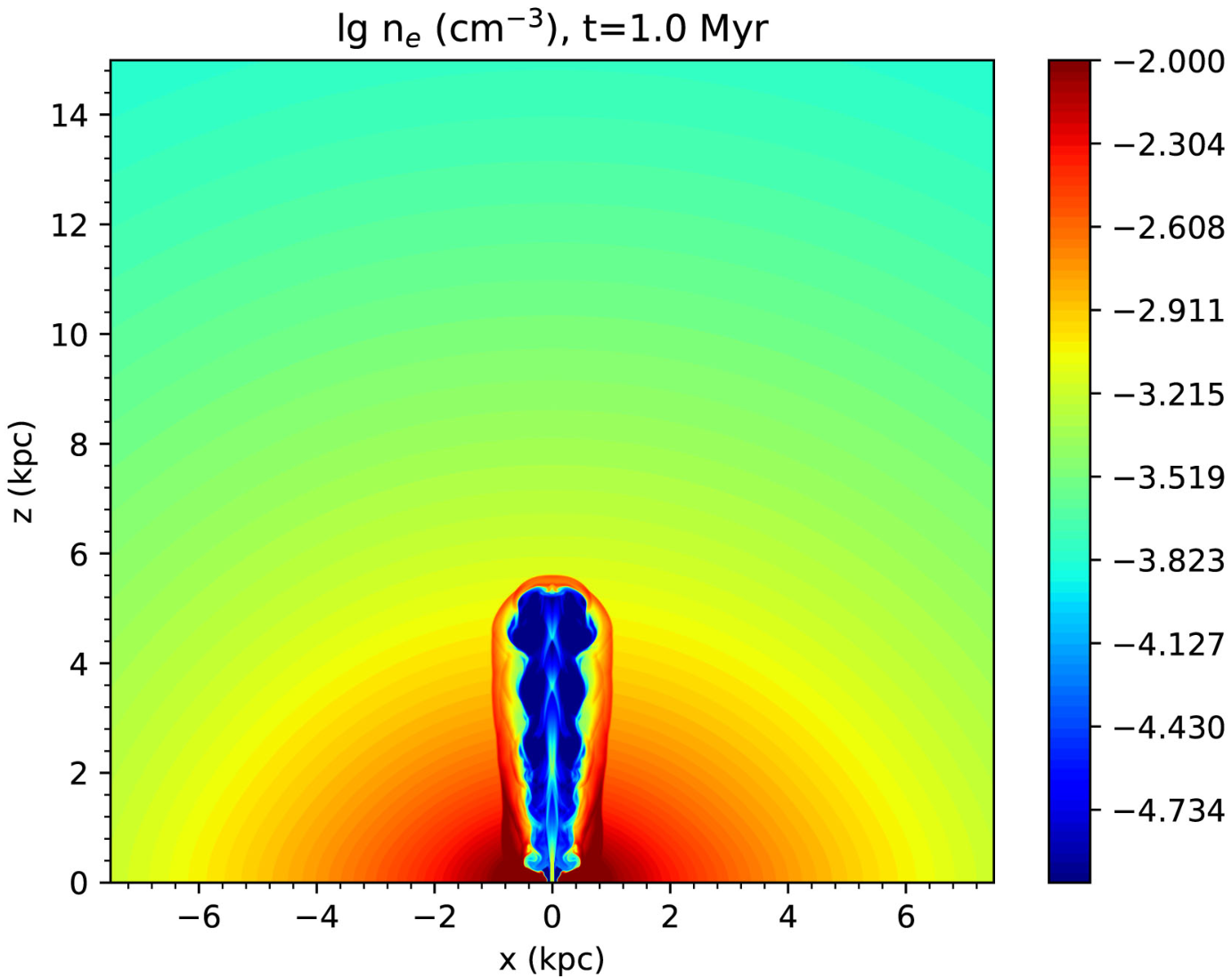}}
\subfigure[]{
   \includegraphics[width=8cm]{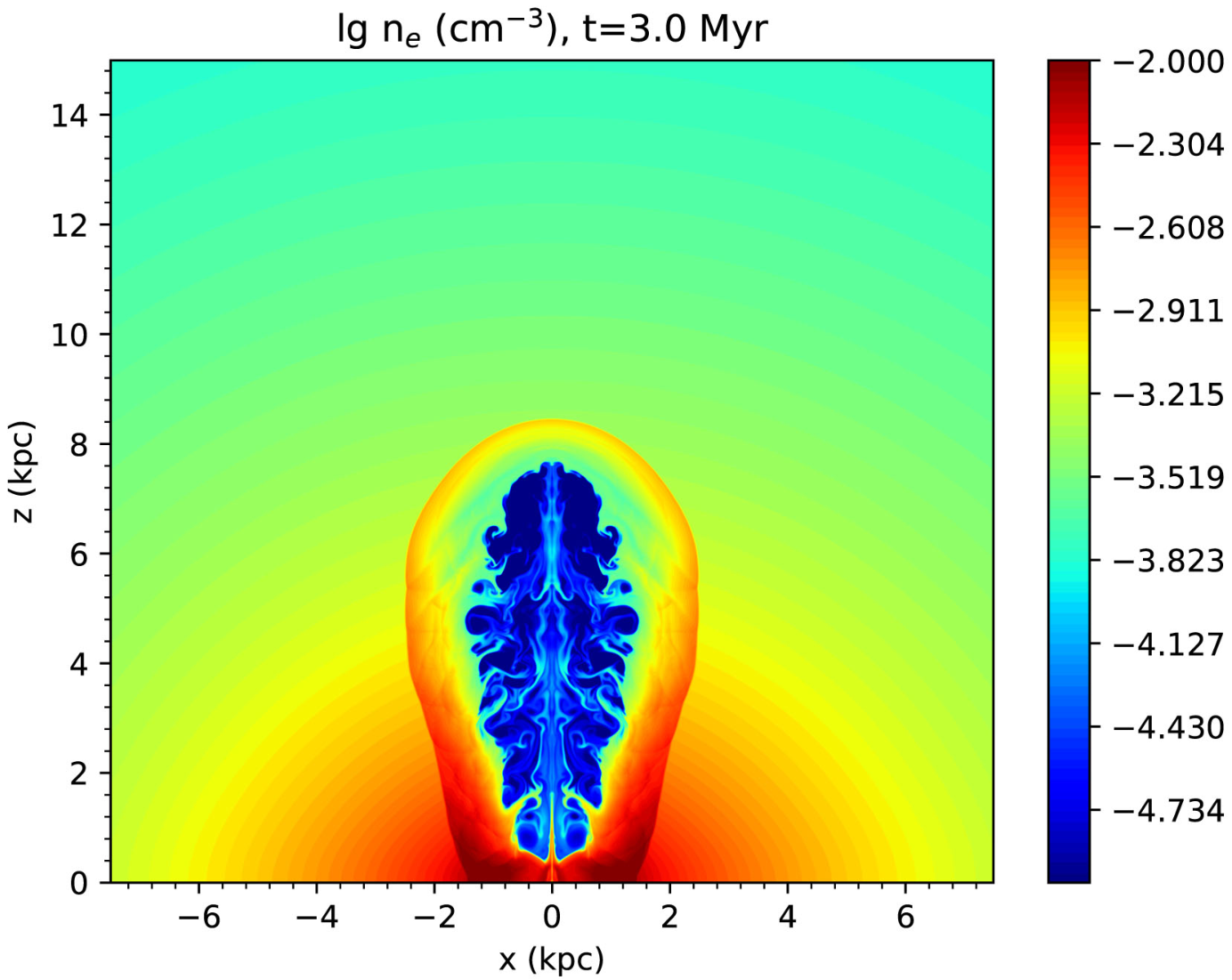}}
 \subfigure[]{
   \includegraphics[width=8cm]{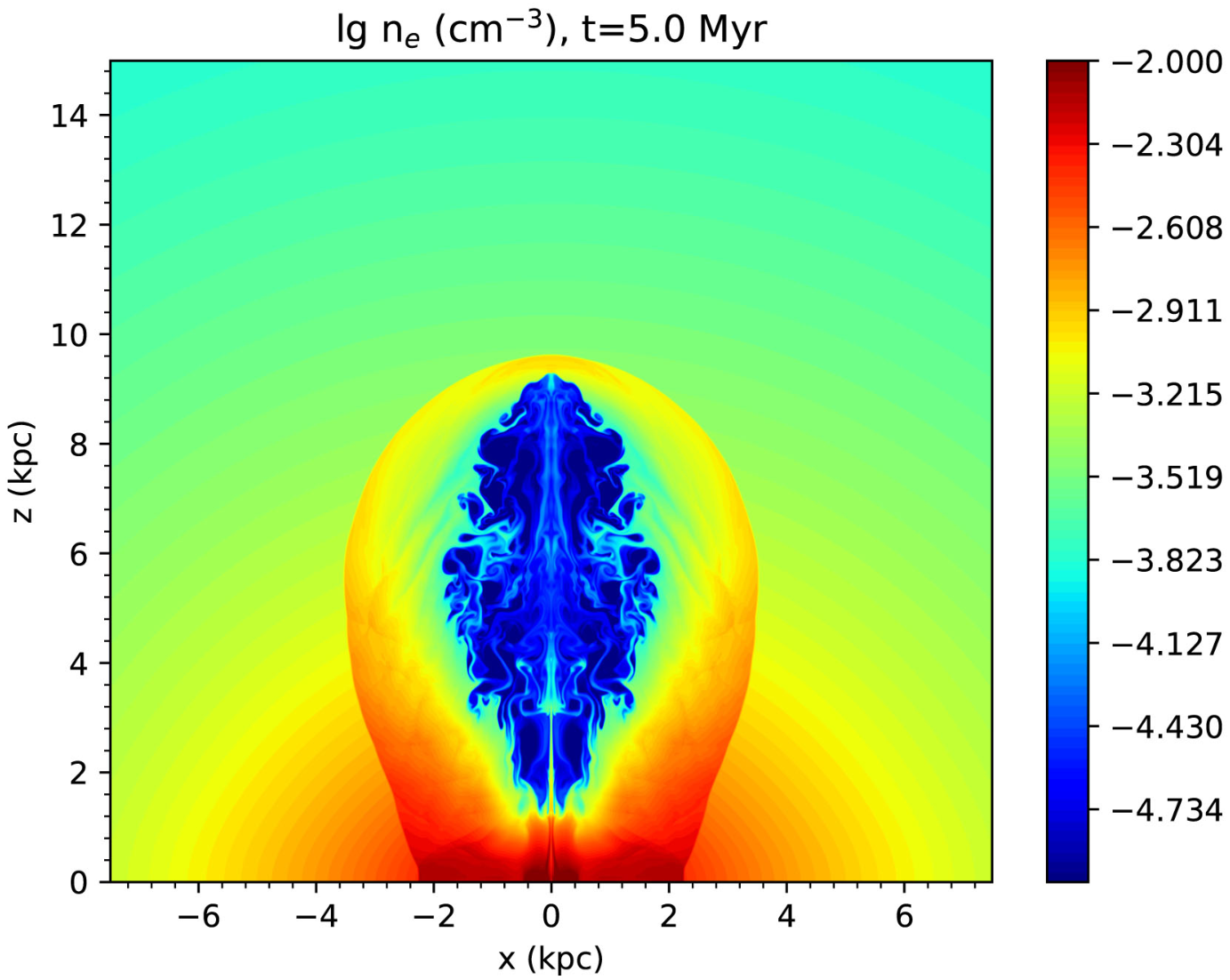}}
\caption{Central slices ($16 \times15$   kpc) of thermal gas density in logarithmic scale at t=1, 3, 5 Myr in run A. Note that the edge of the observed Fermi bubbles corresponds to the expanding forward shock in our model, where CRs are expected to be accelerated.}\label{fig-density}
 \end{figure}

The temporal evolution of the simulated Fermi bubble in run A is shown in Figure \ref{fig-density}, which shows the distribution of thermal electron number density at three different times t= 1, 3, 5 Myr. A forward shock is generated as soon as the jet punches through the ambient halo gas. At t = 1 Myr, the jet is switched off, and the height of the shock reaches $z=6$ kpc while the width is still less than 2 kpc. At $t\simeq 5$ Myr, the bubble expands to its current size as observed. In the shock model, the expanding forward shock accelerates CRs and the region enclosed by the forward shock corresponds to the observed Fermi bubble. As seen from Figure \ref{fig-density}c, the bubble may be divided into two regions: the low-density central lobe, which is enclosed by a contact discontinuity and contains high-temperature jet plasma and some jet-entrained halo gas, and the outer shell, which is located between the forward shock and the inner contact discontinuity and contains the shock-heated halo gas.

\begin{figure}[h!]
\centering
   \includegraphics[width=8cm]{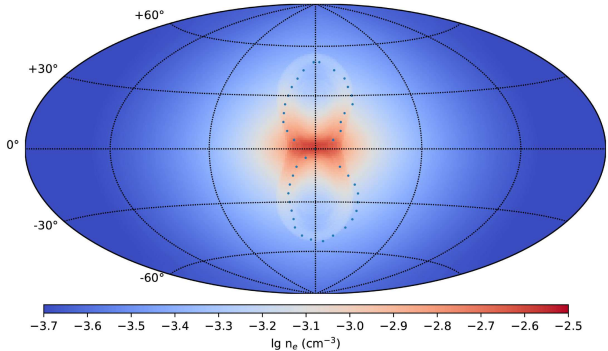}
\caption{Averaged hot gas density along lines of sight from the Earth to a distance of 20 kpc in run A at t = 5 Myr in Galactic coordinates with a Hammer-Aitoff projection. The dots represent the edges of the observed Fermi bubbles. }\label{Figure_density-pro}
 \end{figure}

To directly compare the simulated bubble morphology with observations, we calculate the line-of-sight averaged thermal gas density at t = 5 Myr using Equations \ref{proj} and \ref{proj2}. Figure \ref{Figure_density-pro} shows the averaged thermal electron number density along lines of sight from the Earth to a distance of 20 kpc in Galactic coordinates with a Hammer-Aitoff projection. As seen in this figure, the outline of the projected shock lies quite close to the observed edge of the Fermi bubbles, especially at negative latitudes, suggesting that run A reproduces the location, size and morphology of the Fermi bubbles quite well, and the bubble age in run A is roughly 5 Myr. Our model further predicts that there is a low-density lobe in the middle of each bubble as seen clearly in Figures 4 and 5, and these two low-density lobes may explain the vast cavity of hot gas with radius $\sim 6$ kpc described by \citet{Nicastro2016} in the central region of the Milky Way.

\begin{figure}[h!]
\centering
   \includegraphics[width=8cm]{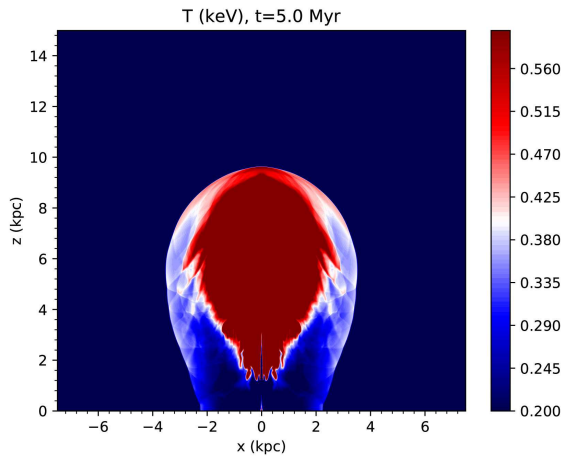}
\caption{Temperature distribution of thermal gas at t = 5 Myr in run A. To better show the temperature distribution in the shock-compressed halo gas, we choose an upper limit of 0.6 keV, which masks high gas temperatures in the central low-density lobe. }\label{fig-temperature}
 \end{figure}

\begin{figure}[h!]
\centering
   \includegraphics[width=8cm]{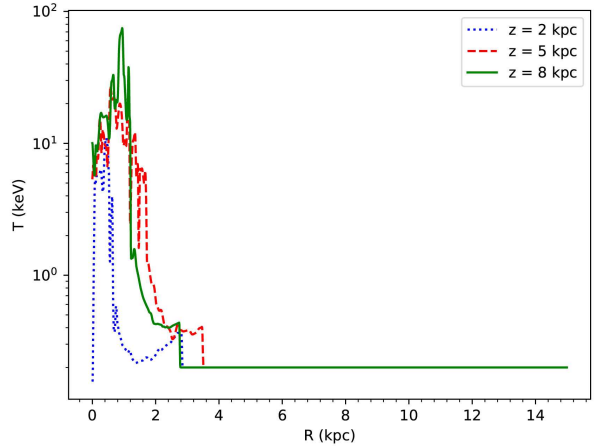}
\caption{Variations of thermal gas temperature along the $R$ direction in run A at t = 5 Myr at three fixed values of $z=2$, $5$, and $8$ kpc. The gas temperatures in the downstream of the forward shock at these three heights are all roughly $0.4$ keV, consistent with those measured by \citet{Miller2016}.}\label{fig-temperature1d}
 \end{figure}

The temperature distribution in the bubble contains very useful information, and by comparing with observations, it can be used to constrain the properties of the Fermi bubbles. Figure \ref{fig-temperature} shows the temperature distribution of thermal gas in run A at t = 5 Myr. The gas temperature in the inner low-density lobe is very high, $\sim 10$ keV or above. In the outer shell, the gas temperature slightly increases from low to high latitudes, and at $z\gtrsim 4$ kpc, the gas temperature is $T\sim 0.4$ keV. This can also be seen in Figure \ref{fig-temperature1d}, which shows the variations of gas temperature along the $R$ direction at t = 5 Myr at three fixed values of $z=2$, $5$, and $8$ kpc. X-ray emission from the inner lobe is expected to be very weak due to the low gas densities there, and there would be essentially no line emissions from this region due to its very high gas temperatures. X-ray emissions from the Fermi bubble would thus be dominated by the outer shell region.

As shown in Figure \ref{fig-temperature1d}, it is remarkable that the gas temperatures in the downstream of the forward shock (the outer shell) at $z=2$, $5$, and $8$ kpc are all quite close to $0.4$ keV, consistent with those measured by \citet{Miller2016} with O IIV and O IIIV emission line ratios. We have also run many additional simulations with different jet powers, and find that the post-shock gas temperature depends quite strongly with the jet power. If the jet is more powerful, it takes less time to form the bubble with the current size, and the post-shock temperature is higher. In this sense, we constrain the age of the Fermi bubbles in the shock model to be $\sim 5$ Myr, and the total energy of this event to be $2P_{\rm jet}t_{j}\sim 2\times 10^{55}$ erg (considering two opposing jets). These estimates are affected by other model parameters, e.g., uncertainties in the ambient halo gas density and the measured gas temperature in the Fermi bubbles, which will be further discussed in Section 4.

\begin{figure}[h!]
 \centering
   \includegraphics[width=8cm]{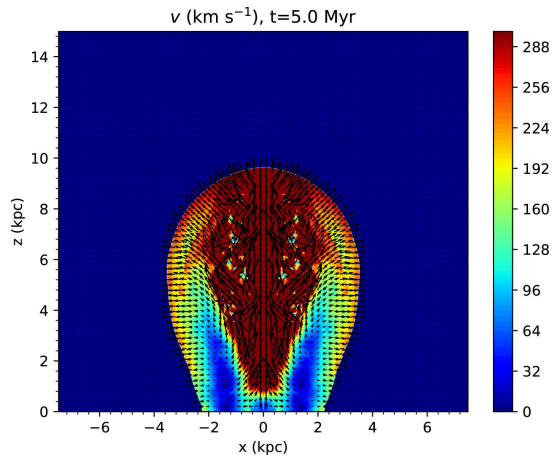}
\caption{Central slices ($16 \times15$ kpc) of thermal gas velocity at t =  5 Myr in run A. Color represents the velocity magnitude, while arrows show its directions. To better show the velocity distribution in the shock-compressed halo gas, we choose an upper limit of 300 km s$^{-1}$, which masks high gas velocities in the central low-density lobe. }\label{fig-vr}
 \end{figure}

Figure \ref{fig-vr} further shows the velocity distribution of thermal gas at t =  5 Myr in run A. While the velocity in the low-density lobe is very high (up to 3000 km s$^{-1}$), the gas velocity in the outer shell is typically $\sim 200-300$ km s$^{-1}$ increasing from low to high latitudes, which is close to the measured velocities of high velocity clouds (HVCs) $91 - 265$ km $s^{-1}$ along several lines of sight towards the Fermi bubbles \citep{Bordoloi2017,Karim2018}.

\subsection{The X-shaped Biconical Structure in X-rays}

The 1.5 keV diffuse X-ray map from the ROSAT all-sky survey revealed an X-shaped biconical structure within 10 degrees around the GC \citep{Snowden1997,bland03,Su2010}, and this inner X-ray structure connects smoothly with the outer Fermi bubbles at latitudes above $10^{\circ}$ \citep{Keshet2017}, suggesting that these two structures share a common origin. Since the forward shock in our model naturally compresses the ambient hot halo gas, the shock model is also expected to explain the X-shaped biconical X-ray structure within 10 degrees around the GC, as shown in this section.

The X-ray surface brightness is calculated for run A at t = 5 Myr as follows. We adopt the APEC plasma model \citep{Smith2001} with a fixed gas metallicity $Z=0.3Z_{\odot}$. Assuming that the hot gas is optically thin and under collisional ionization equilibrium, the surface brightness $I$ in the Galactic coordinates ($l, b$) at the ROSAT 1.5 keV band can be calculated as follows:
\begin{equation}\label{brightness}
I(l,b)=\frac{1}{4\pi}\int_{\rm los} n_{\rm e}^{2}\epsilon(T)dr ~~~ \textrm{erg s}^{-1} \textrm{ cm}^{-2} \textrm{ Sr}^{-1},
\end{equation}
where $\epsilon(T)$ is the volumetric emissivity of the plasma. Atomic data are taken from Astrophysical Plasma Emission Database (APED) with the publicly available PyAtomDB package, and both line emissions and bremsstrahlung are included in $\epsilon(T)$. Along each line of sight, the integration in the above equation is done to a distance of 50 kpc.

\begin{figure}[h!]
\centering
   \includegraphics[width=8cm]{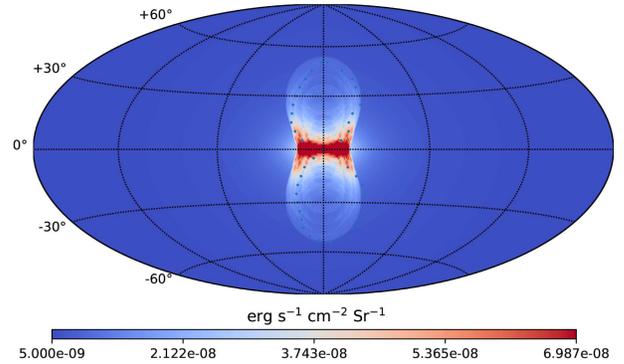}
\caption{Synthetic X-ray (0.7--2 keV) surface brightness map in Galactic coordinates with a Hammer-Aitoff projection for run A at t = 5 Myr. The dots represent the edge of the observed Fermi bubbles.}\label{fig-x-ray}
 \end{figure}

Figure \ref{fig-x-ray} shows the synthetic X-ray (0.7--2 keV) surface brightness map for run A at t = 5 Myr. Due to the compression of hot gas by the forward shock, the simulated Fermi bubble is limb brightened, and in particular, the bubble base is very bright in X-ray, coinciding very well with the location of the bipolar X-ray structure seen in the ROSAT 1.5 keV map. The calculated X-ray surface brightness of the shock-compressed shell at the bubble base is around $5\times10^{-8}$ erg s$^{-1}$ cm$^{-2}$ Sr$^{-1}$, corresponding to $\sim10^{-3}$ counts s$^{-1}$ arcmin$^{-2}$ in the ROSAT R6+R7 band, which is quite close to the observed value of $\sim 5\times10^{-4}$ counts s$^{-1}$ arcmin$^{-2}$. The minor discrepancy could be due to  HI absorption in the Galactic disk and bulge. This result further strengthens the forward shock model for the origin of the Fermi bubbles and the X-shaped biconical structure in the 1.5 keV map.

\subsection{The Emission Measure}

To compare the gas densities in our simulated Fermi bubbles with observations more quantitatively, here we calculate the emission measures (EMs) along the lines of sight toward the bubbles in run A at t = 5 Myr,
\begin{equation}\label{em}
 EM(l,b)=\int_{\rm los}n_{\rm e}^{2}dl ~\textrm{ ,}
\end{equation}
where the integration is done to a distance of 50 kpc from the Earth. We then compare our calculated EMs with the data in \citet{Kataoka2015}, which show the observed EMs along many sight lines toward a very large area of the Fermi bubbles. Using \textit{Suzaku} and \textit{Swift} X-ray data, \citet{Kataoka2015} found that the EM typically decreases with Galactic latitude, varying by an order of magnitude over the region covered by the Fermi bubbles.

Figure \ref{em} shows the EM as a function of Galactic latitude. The orange dots represent the EM data along many sight lines shown in \citet{Kataoka2015}, while the blue dots show the corresponding EMs along the same sight lines calculated in run A. We also calculate the maximum value of the EMs at any given latitude, and show the variation of it with Galactic latitude as the solid blue line. This line represents the EMs along the lines of sight toward the swept-up shell right behind the forward shock. As can be seen, the calculated EM increases from $\sim 0.01 \textrm{~cm}^{-6}~ \textrm{pc}$  at high latitudes to $\sim 0.3  \textrm{~cm}^{-6}~ \textrm{pc}$ near the GC, roughly following the trend in the observations. However, along many sight lines, our calculated EMs are substantially lower than the observed values, which likely include additional contributions from some gaseous structures outside the Fermi bubbles. The asymmetry of the observed EMs between the northern and southern bubbles also suggests that the observed EMs derived from $0.4-10$ keV X-ray observations include significant or even dominant contributions from local structures not directly associated with the Fermi bubble event along many sight lines. Soft X-rays emitted near the GC are subject to strong absorptions, and to probe the gas properties related to the hot Fermi bubbles, it may be better to use hard X-ray observations, such as the biconical X-ray structure revealed by the \textit{ROSAT} 1.5 keV map.

\begin{figure}[h!]
\centering
   \includegraphics[width=8cm]{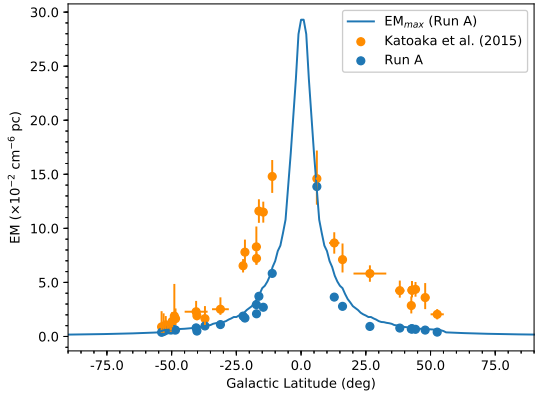}
\caption{\rm Emission measures along many sight lines toward the Fermi bubbles as a function of Galactic latitude. The orange dots represent the EM data in \citet{Kataoka2015}, while the blue dots show the corresponding EMs along the same lines of sights calculated in run A at  t = 5 Myr. We also calculate the maximum value of the EMs toward the sight lines at any given latitude, and show the variation of it with Galactic latitude as the solid blue line, which represents the EMs towards the lines of sight near the forward shock.}\label{em}
 \end{figure}

\subsection{The Mach Number at the Forward Shock}

\begin{figure}[h!]
 \centering
   \includegraphics[width=8cm]{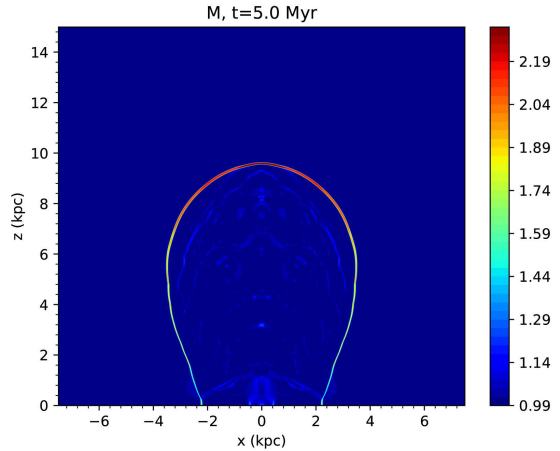}
   \caption{Mach number of the forward shock in Run A  at t = 5 Myr. The Mach number increases from low to high latitudes, with an approximate value of about $M\sim 2$.}\label{fig-Mach1}
    \end{figure}

In the shock model, CRs are accelerated at the expanding forward shock, and diffuse into the bubble interior. The CR acceleration efficiency depends strongly on the Mach number. Here in this subsection we investigate the evolution of the Mach number at the propagating forward shock in our fiducial run (run A). The shock front is identified as jumps in the pressure and temperature distributions and the associated Mach number $M$ is calculated according to the Rankine-Hugoniot jump conditions. At $t =5$ Myr, the Mach number at the forward shock is shown in Figure \ref{fig-Mach1}. As can be seen, the Mach number along the bubble edge increases slightly from $M\sim 1.8$ at $z=2$ kpc to $M\sim 2.2$ at $z=9$ kpc. This value is roughly consistent with the Mach number of $M=2.3^{+1.1}_{-0.4}$ estimated in \citet{Miller2016}. Note that at the bubble top $(R\rm{,} ~z)\sim (0\rm{,} ~9.6 ~\rm{kpc})$ most affected by the jet evolution, the Mach number peaks quickly at $M\sim 2.8$ from nearby regions.

 \begin{figure}[h!]
 \centering
   \includegraphics[width=9cm]{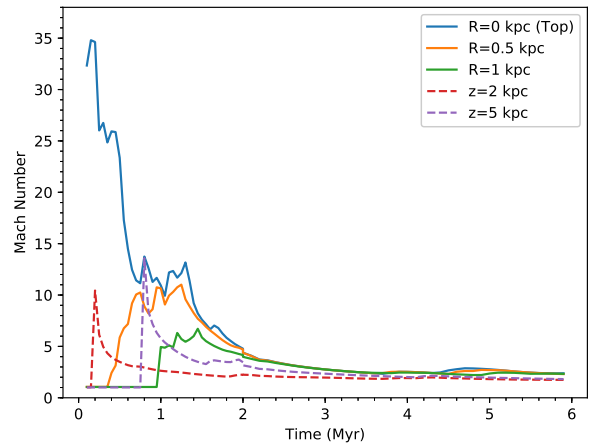}
\caption{Temporal evolution of the Mach number of the forward shock in Run A. From top to bottom, the solid lines refer to the Mach number evolution at $R=0$ (the bubble top), $0.5$ kpc, and $1$ kpc respectively, in the bubble surface. The dashed lines refer to the Mach number evolution at $z=2$ kpc (red), and 5 kpc (purple) in the bubble surface.}\label{fig-Mach2}
 \end{figure}

The temporal evolution of the Mach number at the shock front, i.e. the bubble surface, is shown in Figure \ref{fig-Mach2}. The Mach number evolution at the top of the bubble ($R=0$) clearly shows that a strong forward shock with $M>10$ forms once the jet punches through the ambient halo gas. As the shock front propagates outward, the Mach number at the bubble top roughly decreases from about 30 at t = 0.1 Myr to $\sim 2.8$ at t = 5 Myr. Several fluctuations in the Mach number evolution are caused by the interaction of the jet with gas circulations in the bubble. The three solid lines in Figure \ref{fig-Mach2} show the Mach number evolution at $R=0$ (the bubble top), $0.5$ kpc, and $1$ kpc in the bubble surface, indicating that the Mach number in the head region of the shock front is larger than $\sim 4$ during the first 2 Myrs and drops below $4$ afterwards. The dashed lines correspond to the Mach number evolution at $z=2$ and 5 kpc in the bubble surface, indicating that the Mach number is generally less than $4$ in the middle and bottom regions of the bubble surface. It is generally believed that CR acceleration is inefficient at small Mach numbers. Figure \ref{fig-Mach2} thus shows that CR acceleration is most efficient in the head region of the Fermi bubbles during the early stage of the bubble evolution ($t\lesssim 2$ Myr). Our results here on the Mach number evolution would be useful for future studies of the Fermi bubbles on the CR acceleration processes and the associated non-thermal emissions.

\section{A Parameter Study of the Shock Model}\label{discussion}

In the previous section, we have presented the results of our best-fit fiducial run, which reproduces the location, size, morphology and the gas properties of the Fermi bubbles quite well. The consistency between the simulation and observations indicates that the forward shocks driven by a pair of opposing AGN jets about 5 Myr ago in the GC could possibly be the physical origin of the Fermi bubbles. In this section, we will further explore the shock model of the Fermi bubbles by investigating the uncertainties in the density distribution of the halo gas and the observed gas temperature in the Fermi bubbles. We will also investigate the spherical wind model in the shock scenario of the Fermi bubbles.

\subsection{The Density Distribution of the Halo Gas}

\begin{figure}[h!]
 \centering
   \includegraphics[width=8cm]{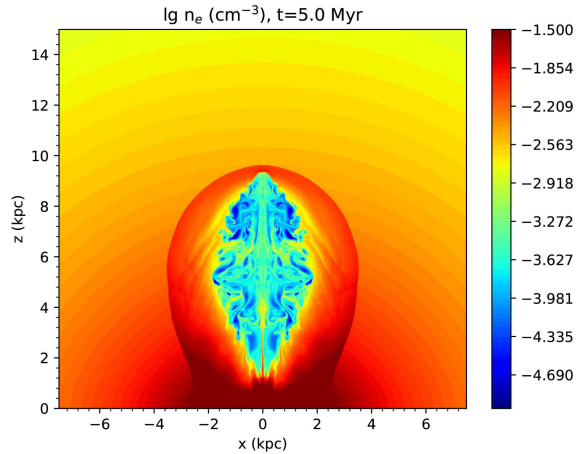}
   \caption{The electron number density distribution of thermal gas at  $t = 5$ Myr in run B, where the initial gas density everywhere in the halo is ten times that in run A. }\label{fig-run4}
    \end{figure}

The normalization of the initial density distribution of the halo gas is determined by the central gas density $n_{e0}$. As shown in \citet{Guo2012a}, the total energy required to produce the observed Fermi bubbles scales linearly with the value of $n_{e0}$. Here we present the results of run B, where $n_{e0}$ is chosen to be $0.3$ cm$^{-3}$, ten times the value in run A. Due to the self-similarity of the hydrodynamic equations (1) - (4), the bubble (forward shock) with the same morphology will be produced if the values of $\rho_{j}$, $e_{j}$, and $e_{cr}$ for the jet setup in run B are also ten times the corresponding values in run A respectively (see Table 1). Figure \ref{fig-run4} shows the gas density distribution at $t = 5$ Myr in run B, which is essentially the same as the bottom panel of Figure \ref{fig-density} for run A at the same time except that the density normalization in Figure \ref{fig-run4} is ten times larger everywhere.

Here we argue that the initial gas density distribution in the halo should be quite close to that in run A and the uncertainty in $n_{e0}$ is small. First, as shown in Figure \ref{fig-miller} the density distribution in run A agrees very well with the $\beta$ model inferred from X-ray observations in \cite{Miller2015}. Second, the O IIV and O IIIV emission line observations towards the Fermi bubbles by \citet{Miller2016} suggest that thermal electron number density in the downstream of the forward shock (the shell region) is about $10^{-3}$ cm$^{-3}$, very close to the corresponding value in run A at $t = 5$ Myr (see Figure \ref{fig-density}). Third, the X-ray surface brightness depends sensitively on the gas density ($I\propto n_{e}^{2}$). If the gas density increases by a factor of 10 as in run B, the X-ray surface brightness of the X-shaped biconical structure near the GC would increase by a factor of 100, contradicting with the ROSAT result which agrees with run A quite well as shown in Sec. 3.2. Thus the initial gas density distribution in the halo is constrained quite well in our hydrostatic model, and would not cause significant uncertainties in the estimations of the total energy and the power of the Fermi bubble event.

\subsection{The Gas Temperature in the Fermi Bubbles }

\begin{figure}[h!]
  \centering
   \includegraphics[width=8cm]{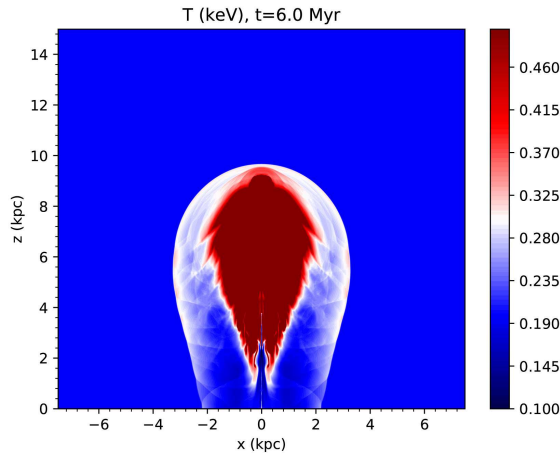}
\caption{The temperature distribution of thermal gas at $t=6$ Myr in run C. The postshock gas temperature in the resulting Fermi bubble is around 0.3 keV, smaller than the value of 0.4 keV in run A. }\label{fig-runc-d}
 \end{figure}

The temperature of the shocked gas in the Fermi bubbles is one of the most important constraints for our model. As discussed in Sec. 3.1, it significantly affects the age and power of the Fermi bubble event. In Run A we adopt the temperature constraint given by \citet{Miller2016}, which is $\sim0.4$ keV for the postshock shell region. In this simulation, it takes 5 Myr for the forward shock to propagate to the current position of the Fermi bubble edge and the postshock gas is heated to $\sim0.4$ keV. Considering the uncertainties in the temperature measurement, here we present the results of an additional run (run C), where the postshock temperature in the bubble is taken to be $0.3$ keV as observed in \citet{Kataoka2015}. To fit both the morphology and the postshock gas temperature of the bubble, we modify several jet parameters in run C as listed in Table \ref{tab-para}. The temperature distribution of thermal gas at $t=6$ Myr in run C is shown in Figure \ref{fig-runc-d}. As seen in this figure, the morphology of the forward shock is very similar to that in the bottom panel of Figuire \ref{fig-density} for run A at $t=5$ Myr, and the postshock gas temperature in the bubble is indeed $\sim0.3$ keV.

Compared to run A, the jet in run C is less powerful and it takes a longer time for the forward shock to propagate to the current location. The jet power in run C is $P_{\rm j}=1.15\times10^{41}$ erg s$^{-1}$. With the jet duration of 1 Myr, the total energy injected by the two opposing jets is $2P_{\rm j}t_{j}=7.22 \times10^{54}$ erg. The current age of the Fermi bubbles in run C is $t_{\rm bub}=6$ Myr. Combining the results of runs A and C, the current age of the Fermi bubbles can be estimated to be $5-6$ Myr, and the total energy of this event is around $(0.7-2.2) \times10^{55}$ erg.

\subsection{The Spherical Wind Model }

In this subsection, we present a preliminary study on the spherical wind model, where the forward shock is driven by starburst or AGN winds at the GC. In previous wind models (e.g., \citealt{Mou2014,Sarkar2017}), the Fermi bubble edge is usually identified as the contact discontinuity. Here in the shock model, we assume that the wind-driven forward shock is the edge of the Fermi bubbles and accelerates CRs diffusing into the bubbles. In wind models, the central molecular zone (CMZ) is often included to suppress the lateral expansion of the bubbles at low latitudes. We adopt a simple setup as in \citet{Sarkar2017}, assuming that the CMZ is a ring-like structure  (e.g., \citealt{Morris1996}) with an inner radius of 80 pc and an outer radius of 240 pc along the Galactic plane, and its height-to-radius ratio is a constant $0.25$.
The gas density in the CMZ is $50 m_{p}$ cm$^{-3}$, where $m_{p}$ is the proton mass, and the CMZ is under local pressure balance with the ambient hot halo gas. Rotation is neglected, and we instead use an artificial centrifugal potential in the CMZ region to counteract the gravity on the CMZ. The wind is initiated as a spherical outflow from the inner 42 pc region at the GC with a constant radial velocity.

 \begin{figure}[h!]
  \centering
   \includegraphics[width=8cm]{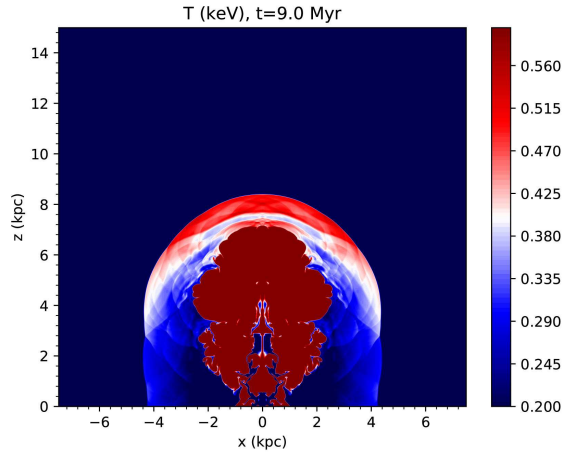}
\caption{The temperature distribution of thermal gas at $t=9$ Myr in run D for the spherical wind model. }\label{fig-rune-d}
 \end{figure}

We performed a large number of wind simulations with different wind parameters (e.g. the wind velocity, density, and duration) and investigated if the spherical wind model can produce a bubble enclosed by the forward shock that meets both the temperature and morphology constraints as described in Section 3. Here we present the results of a representative wind simulation (run D), in which the relevant model parameters are listed in Table \ref{tab-para}. As shown in Figure \ref{fig-rune-d}, the height of the forward shock front at $t=9$ Myr is comparable to that of the observed Fermi bubbles, and the average postshock gas temperature in the bubble is also close to the observed value of 0.4 keV in \citet{Miller2016}. While the lateral expansion of the inner high-temperature lobe at low latitudes is strongly limited by the CMZ as also shown in \citet{Mou2014}, the forward shock can easily bypass the CMZ, producing a bubble with very wide base. Although we tried a large number of wind simulations, the wide base for the forward shock near the Galactic plane is a general feature. This is mainly due to the fact that the height of the CMZ is too small to stop the shock from passing over it. The wide base of the shock front is clearly inconsistent with the narrow base of the observed Fermi bubbles in gamma rays and could not explain the X-shaped biconical X-ray structure near the GC. We thus conclude that starburst or AGN winds are very unlikely to be the origin of the Fermi bubbles in the shock model.

\section{Summary and Discussion}

In this paper, we present a series of hydrodynamic simulations to study the forward shock model for the origin of the Fermi bubbles. We assume that the bubble edge is the forward shock driven by a pair of opposing AGN jets emanating from the GC several Myrs ago and the CRs in the bubbles are mainly accelerated at the forward shock. By properly choosing jet parameters, our model naturally reproduces the observed morphology of the Fermi bubbles, the postshock gas temperature ($0.3 - 0.4$ keV) and the bubble age ($5-6$ Myr) inferred by recent X-ray and ultraviolet observations. Furthermore, the forward shock compresses the hot gas in the halo, and in the predicted X-ray surface brightness map, the compressed gas leads to an X-shaped structure at low latitudes near the GC, which is morphologically very similar to the biconical X-ray structure in the ROSAT 1.5 keV map with very close values of X-ray surface brightness. Thus our shock model simultaneously explains the origins of the Fermi bubbles and the biconical X-ray structure near the GC.

Our simulations indicate that the current age of the Fermi bubbles is $\sim 5-6$ Myr, and the total energy of this event is around $(0.7-2.2) \times10^{55}$ erg. The total energy required to produce the bubbles depends strongly on the halo gas densities which we show are very well constrained by the observed gas density in the shock-swept shell and the observed X-ray surface brightness of the biconical X-ray structure near the GC. In our fiducial run, the two jets last for a duration of 1 Myr with a total power of $2P_{\rm j}\sim 6.84\times10^{41}$ erg s$^{-1}$, corresponding to a mass accretion rate of $\dot{M}_{\rm BH}=2P_{\rm j}/(0.1c^{2})=1.2\times 10^{-4}$ $M_{\sun}/$yr for Sgr A* and an Eddington ratio of $\epsilon \sim1.2\times 10^{-3}$, which falls well in the range of the hot accretion flow mode for SMBHs.

We also experimented with the spherical wind model using a large number of simulations, and found that the base of the resulting bubbles enclosed by the forward shock is generally much wider than observed, suggesting that starburst or AGN winds are very unlikely to be the origin of the Fermi bubbles in the shock model where the bubble edge is a forward shock.

We also present the temporal evolution of the Mach number at the shock front in our simulations, which would be useful for future studies on the CR acceleration process in the Fermi bubbles. Our simulations show that CR acceleration is most efficient in the head regions of the Fermi bubbles during the first 2 Myrs of the bubble evolution. Our simulations are hydrodynamic, and do not follow the evolution of the CR spectrum. Thus, we could not determine if diffusive shock acceleration (DSA) at the shock front is the main acceleration mechanism for the CRs inside the Fermi bubbles and if the gamma ray emissions from the bubbles are mainly hadronic or leptonic. \citet{Fujita2013,Fujita2014} demonstrate that the forward shock model could explain both the observed hard spectrum and the relatively uniform surface brightness distribution of the gamma ray emissions from the bubbles. However, we note that both the total energy and the postshock gas density in their model are much higher than in our simulations, suggesting that further studies on the shock acceleration mechanism of the CRs in the Fermi bubbles are necessary. If DSA alone is not responsible for most CRs in the bubbles, other acceleration mechanisms such as turbulent acceleration \citep{Mertsch2019} in the shock downstream regions may contribute substantially to CR acceleration in the bubbles.

The forward shock model of the Fermi bubbles proposed in this work is intrinsically different compared to previous hydrodynamic models in the literature. Both previous leptonic AGN jet models \citep{Guo2012a,Yang2012} and hadronic wind models \citep{Mou2014,Sarkar2017} assume that the Fermi bubbles are essentially the ejecta bubbles enclosed by the contact discontinuity. In contrast, in our current model the Fermi bubbles are a pair of jet-driven outflowing bubbles enclosed by the forward shock. As seen in Figure 4, the low-density jet ejecta bubbles are located well inside our simulated Fermi bubbles. In previous models, the forward shock front driven by the Fermi bubble event is located much further away from the Fermi bubbles. Therefore, our forward shock model is the first hydrodynamic model that explains the Fermi bubbles and the biconical X-ray structure near the GC as the same phenomenon. Future observations of the forward shock at high latitudes and the non-thermal emissions from the inner ejecta bubbles would put useful constraints on the models of the Fermi bubbles.

\acknowledgments
We thank the anonymous referee for an insightful report. We also thank Joss Bland-Hawthorn for helpful discussions.
This work was supported by the National Natural Science Foundation of China
(Grant Nos. 11873072 and 11633006), the Natural Science Foundation of Shanghai (No. 18ZR1447100),
and the Chinese Academy of Sciences through the Key Research Program of Frontier Sciences (No. QYZDB-SSW-SYS033
and QYZDJ-SSW-SYS008). The calculations presented in this work were performed using the high performance computing resources in the Core Facility for Advanced Research Computing at Shanghai Astronomical Observatory.

\bibliography{ms}

\begin{thebibliography}{}
\expandafter\ifx\csname natexlab\endcsname\relax\def\natexlab#1{#1}\fi
\providecommand{\url}[1]{\href{#1}{#1}}

\bibitem[{{Ackermann} {et~al.}(2014){Ackermann}, {Albert}, {Atwood}, {Baldini},
  {Ballet}, {Barbiellini}, {Bastieri}, {Bellazzini}, {Bissaldi}, {Blandford},
  {Bloom}, {Bottacini}, {Brandt}, {Bregeon}, {Bruel}, {Buehler}, {Buson},
  {Caliandro}, {Cameron}, {Caragiulo}, {Caraveo}, {Cavazzuti}, {Cecchi},
  {Charles}, {Chekhtman}, {Chiang}, {Chiaro}, {Ciprini}, {Claus},
  {Cohen-Tanugi}, {Conrad}, {Cutini}, {D'Ammando}, {de Angelis}, {de Palma},
  {Dermer}, {Digel}, {Di Venere}, {Silva}, {Drell}, {Favuzzi}, {Ferrara},
  {Focke}, {Franckowiak}, {Fukazawa}, {Funk}, {Fusco}, {Gargano}, {Gasparrini},
  {Germani}, {Giglietto}, {Giordano}, {Giroletti}, {Godfrey}, {Gomez-Vargas},
  {Grenier}, {Guiriec}, {Hadasch}, {Harding}, {Hays}, {Hewitt}, {Hou},
  {Jogler}, {J{\'o}hannesson}, {Johnson}, {Johnson}, {Kamae}, {Kataoka},
  {Kn{\"o}dlseder}, {Kocevski}, {Kuss}, {Larsson}, {Latronico}, {Longo},
  {Loparco}, {Lovellette}, {Lubrano}, {Malyshev}, {Manfreda}, {Massaro},
  {Mayer}, {Mazziotta}, {McEnery}, {Michelson}, {Mitthumsiri}, {Mizuno},
  {Monzani}, {Morselli}, {Moskalenko}, {Murgia}, {Nemmen}, {Nuss}, {Ohsugi},
  {Omodei}, {Orienti}, {Orlando}, {Ormes}, {Paneque}, {Panetta}, {Perkins},
  {Pesce-Rollins}, {Petrosian}, {Piron}, {Pivato}, {Rain{\`o}}, {Rando},
  {Razzano}, {Razzaque}, {Reimer}, {Reimer}, {S{\'a}nchez-Conde}, {Schaal},
  {Schulz}, {Sgr{\`o}}, {Siskind}, {Spandre}, {Spinelli}, {Stawarz}, {Strong},
  {Suson}, {Tahara}, {Takahashi}, {Thayer}, {Tibaldo}, {Tinivella}, {Torres},
  {Tosti}, {Troja}, {Uchiyama}, {Vianello}, {Werner}, {Winer}, {Wood}, {Wood},
  \& {Zaharijas}}]{Ackermann2014}
{Ackermann}, M., {Albert}, A., {Atwood}, W.~B., {et~al.} 2014, \apj, 793, 64

\bibitem[{{Axford} {et~al.}(1977){Axford}, {Leer}, \& {Skadron}}]{Axford1977}
{Axford}, W.~I., {Leer}, E., \& {Skadron}, G. 1977, International Cosmic Ray
  Conference, 11, 132

\bibitem[{{Bell}(1978)}]{Bell1978}
{Bell}, A.~R. 1978, \mnras, 182, 147

\bibitem[{{Bland-Hawthorn} \& {Cohen}(2003)}]{bland03}
{Bland-Hawthorn}, J., \& {Cohen}, M. 2003, \apj, 582, 246

\bibitem[{{Bland-Hawthorn} {et~al.}(2019){Bland-Hawthorn}, {Maloney},
  {Sutherland}, {Groves}, {Guglielmo}, {Li}, {Curzons}, {Cecil}, \&
  {Fox}}]{bland19}
{Bland-Hawthorn}, J., {Maloney}, P.~R., {Sutherland}, R., {et~al.} 2019, \apj,
  886, 45

\bibitem[{{Blandford} \& {Ostriker}(1978)}]{Blandford1978}
{Blandford}, R.~D., \& {Ostriker}, J.~P. 1978, \apjl, 221, L29

\bibitem[{{Bordoloi} {et~al.}(2017){Bordoloi}, {Fox}, {Lockman}, {Wakker},
  {Jenkins}, {Savage}, {Hernandez}, {Tumlinson}, {Bland-Hawthorn}, \&
  {Kim}}]{Bordoloi2017}
{Bordoloi}, R., {Fox}, A.~J., {Lockman}, F.~J., {et~al.} 2017, \apj, 834, 191

\bibitem[{{Bregman} {et~al.}(2018){Bregman}, {Anderson}, {Miller},
  {Hodges-Kluck}, {Dai}, {Li}, {Li}, \& {Qu}}]{Bregman2018}
{Bregman}, J.~N., {Anderson}, M.~E., {Miller}, M.~J., {et~al.} 2018, \apj, 862,
  3

\bibitem[{{Cheng} {et~al.}(2014){Cheng}, {Chernyshov}, {Dogiel}, \&
  {Ko}}]{Cheng2014}
{Cheng}, K.~S., {Chernyshov}, D.~O., {Dogiel}, V.~A., \& {Ko}, C.~M. 2014,
  \apj, 790, 23

\bibitem[{{Cheng} {et~al.}(2015){Cheng}, {Chernyshov}, {Dogiel}, \&
  {Ko}}]{Cheng2015}
---. 2015, \apj, 804, 135

\bibitem[{{Cheng} {et~al.}(2011){Cheng}, {Chernyshov}, {Dogiel}, {Ko}, \&
  {Ip}}]{Cheng2011}
{Cheng}, K.~S., {Chernyshov}, D.~O., {Dogiel}, V.~A., {Ko}, C.~M., \& {Ip},
  W.~H. 2011, \apjl, 731, L17

\bibitem[{{Crocker} \& {Aharonian}(2011)}]{Crocker2011}
{Crocker}, R.~M., \& {Aharonian}, F. 2011, Physical Review Letters, 106, 101102

\bibitem[{{Crocker} {et~al.}(2015){Crocker}, {Bicknell}, {Taylor}, \&
  {Carretti}}]{Crocker2015}
{Crocker}, R.~M., {Bicknell}, G.~V., {Taylor}, A.~M., \& {Carretti}, E. 2015,
  \apj, 808, 107

\bibitem[{{Dobler} {et~al.}(2010){Dobler}, {Finkbeiner}, {Cholis}, {Slatyer},
  \& {Weiner}}]{dobler10}
{Dobler}, G., {Finkbeiner}, D.~P., {Cholis}, I., {Slatyer}, T., \& {Weiner}, N.
  2010, \apj, 717, 825

\bibitem[{{Fang} {et~al.}(2013){Fang}, {Bullock}, \&
  {Boylan-Kolchin}}]{Fang2013}
{Fang}, T., {Bullock}, J., \& {Boylan-Kolchin}, M. 2013, \apj, 762, 20

\bibitem[{{Fang} {et~al.}(2020){Fang}, {Guo}, \& {Yuan}}]{fang20}
{Fang}, X.-E., {Guo}, F., \& {Yuan}, Y.-F. 2020, arXiv e-prints,
  arXiv:2001.10836

\bibitem[{{Fermi}(1949)}]{Fermi1949}
{Fermi}, E. 1949, Il Nuovo Cimento, 6, 317

\bibitem[{{Finkbeiner}(2004)}]{finkbeiner04a}
{Finkbeiner}, D.~P. 2004, \apj, 614, 186

\bibitem[{{Fox} {et~al.}(2015){Fox}, {Bordoloi}, {Savage}, {Lockman},
  {Jenkins}, {Wakker}, {Bland-Hawthorn}, {Hernandez}, {Kim}, {Benjamin},
  {Bowen}, \& {Tumlinson}}]{fox15}
{Fox}, A.~J., {Bordoloi}, R., {Savage}, B.~D., {et~al.} 2015, \apjl, 799, L7

\bibitem[{{Fujita} {et~al.}(2013){Fujita}, {Ohira}, \& {Yamazaki}}]{Fujita2013}
{Fujita}, Y., {Ohira}, Y., \& {Yamazaki}, R. 2013, \apjl, 775, L20

\bibitem[{{Fujita} {et~al.}(2014){Fujita}, {Ohira}, \& {Yamazaki}}]{Fujita2014}
---. 2014, \apj, 789, 67

\bibitem[{{Grcevich} \& {Putman}(2009)}]{Grcevich2009}
{Grcevich}, J., \& {Putman}, M.~E. 2009, \apj, 696, 385

\bibitem[{{Guo}(2015)}]{guo15}
{Guo}, F. 2015, \apj, 803, 48

\bibitem[{{Guo}(2017)}]{guo17}
{Guo}, F. 2017, in IAU Symposium, Vol. 322, The Multi-Messenger Astrophysics of
  the Galactic Centre, ed. R.~M. {Crocker}, S.~N. {Longmore}, \& G.~V.
  {Bicknell}, 189--192

\bibitem[{{Guo} {et~al.}(2018){Guo}, {Duan}, \& {Yuan}}]{guo18}
{Guo}, F., {Duan}, X., \& {Yuan}, Y.-F. 2018, \mnras, 473, 1332

\bibitem[{{Guo} \& {Mathews}(2011)}]{guo11}
{Guo}, F., \& {Mathews}, W.~G. 2011, \apj, 728, 121

\bibitem[{{Guo} \& {Mathews}(2012)}]{Guo2012a}
---. 2012, \apj, 756, 181

\bibitem[{{Guo} {et~al.}(2012){Guo}, {Mathews}, {Dobler}, \& {Oh}}]{Guo2012}
{Guo}, F., {Mathews}, W.~G., {Dobler}, G., \& {Oh}, S.~P. 2012, \apj, 756, 182

\bibitem[{{Henley} \& {Shelton}(2013)}]{Henley2013}
{Henley}, D.~B., \& {Shelton}, R.~L. 2013, \apj, 773, 92

\bibitem[{{Heywood} {et~al.}(2019){Heywood}, {Camilo}, {Cotton}, {Yusef-Zadeh},
  {Abbott}, {Adam}, {Aldera}, {Bauermeister}, {Booth}, {Botha}, {Botha},
  {Brederode}, {Brits}, {Buchner}, {Burger}, {Chalmers}, {Cheetham}, {de
  Villiers}, {Dikgale-Mahlakoana}, {du Toit}, {Esterhuyse}, {Fanaroff},
  {Foley}, {Fourie}, {Gamatham}, {Goedhart}, {Gounden}, {Hlakola}, {Hoek},
  {Hokwana}, {Horn}, {Horrell}, {Hugo}, {Isaacson}, {Jonas}, {Jordaan},
  {Joubert}, {J{\'o}zsa}, {Julie}, {Kapp}, {Kenyon}, {Kotz{\'e}}, {Kriel},
  {Kusel}, {Lehmensiek}, {Liebenberg}, {Loots}, {Lord}, {Lunsky}, {Macfarlane},
  {Magnus}, {Magozore}, {Mahgoub}, {Main}, {Malan}, {Malgas}, {Manley},
  {Maree}, {Merry}, {Millenaar}, {Mnyandu}, {Moeng}, {Monama}, {Mphego}, {New},
  {Ngcebetsha}, {Oozeer}, {Otto}, {Passmoor}, {Patel}, {Peens-Hough},
  {Perkins}, {Ratcliffe}, {Renil}, {Rust}, {Salie}, {Schwardt}, {Serylak},
  {Siebrits}, {Sirothia}, {Smirnov}, {Sofeya}, {Swart}, {Tasse}, {Taylor},
  {Theron}, {Thorat}, {Tiplady}, {Tshongweni}, {van Balla}, {van der Byl}, {van
  der Merwe}, {van Dyk}, {Van Rooyen}, {Van Tonder}, {Van Wyk}, {Wallace},
  {Welz}, \& {Williams}}]{Heywood2019}
{Heywood}, I., {Camilo}, F., {Cotton}, W.~D., {et~al.} 2019, \nat, 573, 235

\bibitem[{{Karim} {et~al.}(2018){Karim}, {Fox}, {Jenkins}, {Bordoloi},
  {Wakker}, {Savage}, {Lockman}, {Crawford}, {Jorgenson}, \&
  {Bland-Hawthorn}}]{Karim2018}
{Karim}, M.~T., {Fox}, A.~J., {Jenkins}, E.~B., {et~al.} 2018, \apj, 860, 98

\bibitem[{{Kataoka} {et~al.}(2018){Kataoka}, {Sofue}, {Inoue}, {Akita},
  {Nakashima}, \& {Totani}}]{Kataoka2018}
{Kataoka}, J., {Sofue}, Y., {Inoue}, Y., {et~al.} 2018, Galaxies, 6, 27

\bibitem[{{Kataoka} {et~al.}(2015){Kataoka}, {Tahara}, {Totani}, {Sofue},
  {Inoue}, {Nakashima}, \& {Cheung}}]{Kataoka2015}
{Kataoka}, J., {Tahara}, M., {Totani}, T., {et~al.} 2015, \apj, 807, 77

\bibitem[{{Kataoka} {et~al.}(2013){Kataoka}, {Tahara}, {Totani}, {Sofue},
  {Stawarz}, {Takahashi}, {Takeuchi}, {Tsunemi}, {Kimura}, {Takei}, {Cheung},
  {Inoue}, \& {Nakamori}}]{Kataoka2013}
---. 2013, \apj, 779, 57

\bibitem[{{Keshet} \& {Gurwich}(2017)}]{Keshet2017}
{Keshet}, U., \& {Gurwich}, I. 2017, \apj, 840, 7

\bibitem[{{Krymskii}(1977)}]{Krymskii1977}
{Krymskii}, G.~F. 1977, Akademiia Nauk SSSR Doklady, 234, 1306

\bibitem[{{Lacki}(2014)}]{Lacki2014}
{Lacki}, B.~C. 2014, \mnras, 444, L39

\bibitem[{{Maller} \& {Bullock}(2004)}]{Maller2004}
{Maller}, A.~H., \& {Bullock}, J.~S. 2004, \mnras, 355, 694

\bibitem[{{McMillan}(2011)}]{McMillan2011}
{McMillan}, P.~J. 2011, \mnras, 414, 2446

\bibitem[{{McMillan}(2017)}]{McMillan2017}
---. 2017, \mnras, 465, 76

\bibitem[{{Mertsch} \& {Petrosian}(2019)}]{Mertsch2019}
{Mertsch}, P., \& {Petrosian}, V. 2019, \aap, 622, A203

\bibitem[{Mertsch \& Sarkar(2011)}]{Mertsch2011}
Mertsch, P., \& Sarkar, S. 2011, Physical Review Letters, 107, 1

\bibitem[{{Miller} \& {Bregman}(2015)}]{Miller2015}
{Miller}, M.~J., \& {Bregman}, J.~N. 2015, \apj, 800, 14

\bibitem[{{Miller} \& {Bregman}(2016)}]{Miller2016}
---. 2016, \apj, 829, 9

\bibitem[{{Morris} \& {Serabyn}(1996)}]{Morris1996}
{Morris}, M., \& {Serabyn}, E. 1996, \araa, 34, 645

\bibitem[{{Mou} {et~al.}(2014){Mou}, {Yuan}, {Bu}, {Sun}, \& {Su}}]{Mou2014}
{Mou}, G., {Yuan}, F., {Bu}, D., {Sun}, M., \& {Su}, M. 2014, \apj, 790, 109

\bibitem[{{Mou} {et~al.}(2015){Mou}, {Yuan}, {Gan}, \& {Sun}}]{Mou2015}
{Mou}, G., {Yuan}, F., {Gan}, Z., \& {Sun}, M. 2015, \apj, 811, 37

\bibitem[{{Nakashima} {et~al.}(2018){Nakashima}, {Inoue}, {Yamasaki}, {Sofue},
  {Kataoka}, \& {Sakai}}]{Nakashima2018}
{Nakashima}, S., {Inoue}, Y., {Yamasaki}, N., {et~al.} 2018, \apj, 862, 34

\bibitem[{{Navarro} {et~al.}(1996){Navarro}, {Frenk}, \& {White}}]{Navarro1996}
{Navarro}, J.~F., {Frenk}, C.~S., \& {White}, S.~D.~M. 1996, \apj, 462, 563

\bibitem[{{Nicastro} {et~al.}(2016){Nicastro}, {Senatore}, {Krongold},
  {Mathur}, \& {Elvis}}]{Nicastro2016}
{Nicastro}, F., {Senatore}, F., {Krongold}, Y., {Mathur}, S., \& {Elvis}, M.
  2016, \apjl, 828, L12

\bibitem[{{Sarkar} {et~al.}(2015){Sarkar}, {Nath}, \& {Sharma}}]{Sarkar2015}
{Sarkar}, K.~C., {Nath}, B.~B., \& {Sharma}, P. 2015, \mnras, 453, 3827

\bibitem[{{Sarkar} {et~al.}(2017){Sarkar}, {Nath}, \& {Sharma}}]{Sarkar2017}
---. 2017, \mnras, 467, 3544

\bibitem[{{Sasaki} {et~al.}(2015){Sasaki}, {Asano}, \& {Terasawa}}]{Sasaki2015}
{Sasaki}, K., {Asano}, K., \& {Terasawa}, T. 2015, \apj, 814, 93

\bibitem[{{Smith} {et~al.}(2001){Smith}, {Brickhouse}, {Liedahl}, \&
  {Raymond}}]{Smith2001}
{Smith}, R.~K., {Brickhouse}, N.~S., {Liedahl}, D.~A., \& {Raymond}, J.~C.
  2001, \apjl, 556, L91

\bibitem[{{Snowden} {et~al.}(1997){Snowden}, {Egger}, {Freyberg}, {McCammon},
  {Plucinsky}, {Sanders}, {Schmitt}, {Tr{\"u}mper}, \& {Voges}}]{Snowden1997}
{Snowden}, S.~L., {Egger}, R., {Freyberg}, M.~J., {et~al.} 1997, \apj, 485, 125

\bibitem[{{Sofue}(2019)}]{Sofue2019}
{Sofue}, Y. 2019, \mnras, 484, 2954

\bibitem[{{Stone} \& {Norman}(1992)}]{stone1992}
{Stone}, J.~M., \& {Norman}, M.~L. 1992, \apjs, 80, 753

\bibitem[{{Su} \& {Finkbeiner}(2012)}]{Su2012}
{Su}, M., \& {Finkbeiner}, D.~P. 2012, \apj, 753, 61

\bibitem[{{Su} {et~al.}(2010){Su}, {Slatyer}, \& {Finkbeiner}}]{Su2010}
{Su}, M., {Slatyer}, T.~R., \& {Finkbeiner}, D.~P. 2010, \apj, 724, 1044

\bibitem[{{Yang} {et~al.}(2018){Yang}, {Ruszkowski}, \& {Zweibel}}]{Yang2018}
{Yang}, H.-Y., {Ruszkowski}, M., \& {Zweibel}, E. 2018, Galaxies, 6, 29

\bibitem[{{Yang} \& {Ruszkowski}(2017)}]{Yang2017}
{Yang}, H.-Y.~K., \& {Ruszkowski}, M. 2017, \apj, 850, 2

\bibitem[{{Yang} {et~al.}(2012){Yang}, {Ruszkowski}, {Ricker}, {Zweibel}, \&
  {Lee}}]{Yang2012}
{Yang}, H.-Y.~K., {Ruszkowski}, M., {Ricker}, P.~M., {Zweibel}, E., \& {Lee},
  D. 2012, \apj, 761, 185

\bibitem[{{Yang} {et~al.}(2013){Yang}, {Ruszkowski}, \& {Zweibel}}]{yang13}
{Yang}, H. Y.~K., {Ruszkowski}, M., \& {Zweibel}, E. 2013, \mnras, 436, 2734

\bibitem[{{Yao} {et~al.}(2009){Yao}, {Wang}, {Hagihara}, {Mitsuda}, {McCammon},
  \& {Yamasaki}}]{Yao2009}
{Yao}, Y., {Wang}, Q.~D., {Hagihara}, T., {et~al.} 2009, \apj, 690, 143

\bibitem[{{Yuan} \& {Narayan}(2014)}]{Yuan2014}
{Yuan}, F., \& {Narayan}, R. 2014, \araa, 52, 529

\bibitem[{{Zubovas} {et~al.}(2011){Zubovas}, {King}, \&
  {Nayakshin}}]{Zubovas2011}
{Zubovas}, K., {King}, A.~R., \& {Nayakshin}, S. 2011, \mnras, 415, L21

\end{thebibliography}
\end{document}